\documentclass[11pt]{article}
\usepackage{amsmath} 
\usepackage[dvips]{graphics}
\usepackage{latexsym,psfig}
\usepackage{amscd}     \usepackage{amsxtra}
\usepackage{upref}     \usepackage{amsthm}
\usepackage{amssymb}   
\usepackage{amsfonts}
\begin{document}

\title{{\bf
 A nonlinear relativistic approach to mathematical representation of
vacuum electromagnetism based on extended Lie derivative}}

\author{{\bf Stoil Donev}\footnote{e-mail:sdonev@inrne.bas.bg},
{\bf Maria Tashkova}, \\
Institute for Nuclear
Research and Nuclear Energy,\\ Bulg.Acad.Sci., 1784 Sofia,
blvd.Tzarigradsko chaussee 72\\ Bulgaria\\}
\date{}
\maketitle

\begin{abstract}
This paper presents an alternative {\it relativistic nonlinear} approach to the
vacuum case of classical electrodynamics. Our view is based on the
understanding that the corresponding differential equations should be dynamical
in nature. So, they must represent local energy-momentum balance relations.
Formally, the new equations are in terms of appropriately extended Lie
derivative of $\mathbb{R}^2$-valued differential 2-form along a
$\mathbb{R}^2$-valued 2-vector on Minkowski space-time.
 \end{abstract}

\section{Introduction}
In [1] we presented an alternative view and developed corresponding formal
prerelativistic approach to description of time dependent electromagnetic
fields in vacuum. Our approach was based on the general view that no point-like
and spatially infinite physical objects may exist at all, so the point-like
directed idealization of Coulomb law and the D'Alembert wave equation directed
idealization of vacuum fields we put on reconsideration. Such a view on
physical objects, in particular, on free electromagnetic ones, as {\it
spatially finite entities with internal dynamical structure}, made us try to
work out a new look at the problem of choosing appropriate mathematical images
for their physical and dynamical appearances, in general, and for their time
stable and recognizable subsystems, in particular, during their existence. In
other words, accepting the view for available internal local dynamical
structure of free electromagnetic objects, to try to understand in what way and
in what extent this internal dynamical structure determines the behavior of
these objects as a whole under appropriate invironment. The assumed by us
viewpoint there could be shortly characterised in prerelativestic terms as
follows.

Time-stability of a free electromagnetic object requires at least two
interacting subsystems, and these two subsystems can NOT be formally identified
by the electric and magnetic fields $(\mathbf{E},\mathbf{B})$, for example: the
first - by $\mathbf{E}$, and the second - by $\mathbf{B}$, because {\it a
recognizable subsystem of a propagating electromagnetic physical object must be
able to carry momentum}, and neither $\mathbf{E}$ nor $\mathbf{B}$ are able to
do this separately: the local momentum is $\frac1c\mathbf{E}\times\mathbf{B}$.
The supposed two subsystems might be formally identified in terms of the very
$(\mathbf{E},\mathbf{B})$ and, if needed, making use of their derivatives. In
order to come to a more adequate formal representatives of these two subsystems
we paid due respect to the object that represents formally the physical nature
and appearence of an electromagnetic object. For such mathematical object we
chose the corresponding Maxwell stress tensor $M(\mathbf{E},\mathbf{B})$. Such
a choice was based on the general view that the surviving flexability of any
physical object is determined by its capabilities to act upon and to withstand
external action upon, so the corresponding theoretical quantities describing
the corresponding abilities in our case must have the sense of {\it admissible
changes} of $M(\mathbf{E},\mathbf{B})$. This understanding directed our
attention to the well known formal differential identity satisfied by
$M(\mathbf{E},\mathbf{B})$, and brought us to the mathematical images of the
two recognizable subsystems, to the differential relations describing their
time stability, and to the special interaction respect these two subsystems pay
to each other.

Going back to the subject of the present paper we note that after the deep
studies of Lorentz [2], Poincare [3] and Einstein [4], the final step towards
{\it formal relativisation of Maxwell electrodynamical equations} we due to
H.Minkowski [5], who explicitly introduced the new mathematics: local view on
the space $M=(\mathbb{R}^{4},\eta)$ where $\eta$ is pseudometric tensor field,
the classical electric and magnetic vector fields as constituents of a
differential 2-form $F$, Maxwell equations in terms of $F$ on $M$, appropriate
4-dimensional extension of Maxwell stress tensor, unifying in this way the
concepts of stress, energy and momentum as components af a symmetric 2-tensor
field on $M$, and representing the corresponding local conservation laws as
zero-divergence of this tensor field.

Later on it was found that Maxwell equations presuppose in fact two
differential 2-forms, $(F,*F)$, where $*$ is the Hodge star operator
defined by the pseudometric $\eta$ on $\mathbb{R}^4$. Moreover, the
prerelativistic Maxwell system of equations in the vacuum case was represented
in terms of the exterior derivative $\mathbf{d}$ in the form: $\mathbf{d}F=0,
\mathbf{d}*F=0$. This form of the equations stabilized strongly during the
following years the belief in the 4-potential guage view: the basic
mathematical representative of the field should be $u(1)$-valued 1-form
$A=A_\mu dx^\mu\otimes e$ defined on the Minkowski space-time, where $e$
denotes a basis of the 1-dimensional Lie algebra $u(1)$. This view suggested to
consider the 2-form $F$ as $\mathbf{d}A$, to call it {\it field strength}, but
it also reduced the equation $\mathbf{d}F=0$ to trivial and non - informative
one. As for the other 2-form  $*F$, its non-guage originated differential
$\mathbf{d}*F$ was kept in order to be equalized to the electric current 3-form
$*j$: $\mathbf{d}*F=*j$. This final relation we can not admit as
sufficiently realistic, since mathematics requires on both sides of $"="$ to
stay the {\it same} quantity, and theoretical physics could hardly present a
{\it physical quantity} that could be equally well presented by $\mathbf{d}*F$
and by $*j$ in view of their {\it quite different qualitative nature}.
Nevertheless, this formal view clearly suggests that,
from physical point of view, the electromagnetic field should be considered as
built of two interacting and recognizable subsystems formally represented by $F$
and $*F$. However, such a view, somehow, was not adopted and further
elaborated.

We must note that this relativistic formulation {\it does
not introduce new solutions}. Moreover, the new form of the equations and the
new formal identity satisfied by the relativistically extended Maxwell stress
tensor, which was called stress-energy-momentum tensor $Q_\mu^\nu$, do not
introduce explicitly anything about possible nonzero local energy-momentum
exchange between the above mentioned two subsystems formally represented by $F$
and $*F$ (see the next section). In this sense, the essential moments of the
old viewpoint on the field dynamics were kept unchanged, a serious physical
interpretation of $F$ and $*F$ as two physically interacting time-recognizable
subsystems, guaranteing time-stability of an electromagnetic field object
considered as spatially propagating and spatially finite entity carrying
dynamical structure, was not given. In our view, such interpretation is still
not sufficiently well understood today and considered as necessary, also, the
null field nature of the stress-energy-momentum tensor $Q_\mu^\nu:
Q_{\mu\nu}Q^{\mu\nu}=0$, according to us, is not appropriately appreciated,
correspondingly respected, and effectively used.

In this paper we give corresponding to our view relativisation of [1] making
use of modern differential geometry and extending appropriately the Lie
derivative of differential forms along multivector fields [12] to derivative of
vector valued differential forms $\Phi=\Phi^j\otimes k_j\in\Lambda(M,E_1)$
along vector valued multivector fields $T=T_i\otimes e^i\in\mathfrak{X}(M,E_2)$
with respect to appropriately defined bilinear map $\varphi:E_1\times
E_2\rightarrow F$. The role of $\varphi$ is to recognize and partially evaluate
{\it algebraically} and {\it differentially} the time-stable
subsystems represented formally by the components $T_i$ of $T$ and
$\Phi_j$ of $\Phi$, and to separate corresponding
interacting couples $(\Phi^j,T_i)$.

\section{Our basic views}
We start with some general remarks.

First, under {\it physical object/system}, we understand {\it a
system $\mathcal{A}$ of recognizable and spatially finite mutually supporting
subsystems through energy exchange physical processes}, some of these processes
are among the subsystems of $\mathcal{A}$, called by us {\it internal}
with respect to $\mathcal{A}$, e.g., energy exchange between $F$ and
$*F$ in electrodynamics, otherwise, we call them {\it external}
with respect to $\mathcal{A}$, e.g., electric/gravitational field of a
charged/mass particle .

Second, the frequently used in
literature concept of {\it physical system in vacuum}, we understand as {\it
physical system in appropriate media and available appropriate
mutual interaction between the object and the media}, quaranteeng
corresponding time stability and admissible changes of both, the object and the
media.

In view of the above, for electromagnetic field objects we assume:

	{\bf 1.} Every electromagnetic field object exists through {\it
permanent propagation in space with the velocity of light}.

	{\bf 2.} Every electromagnetic field object is built of two {\it field
subsystems}.

	{\bf 3.} These two subsystems stay {\it recognizable} during the entire
existence of the object.

	{\bf 4.} These two subsystems {\it appropriately interect} with each
other, i.e., both are able to carry and to {\it appropriately gain and lose
energy-momentum}.

	{\bf 5.} These two subsystems withstand nonzero {\it recognizable} local
changes coming from the mutual local energy-momentum exchange, so we are going
to consider these changes as {\it admissible}.

The obove views say that we consider electromagnetic field objects as {\it
real, massless, time-stable and space propagating physical objects with
intrinsically compatible and time-stable dynamical structure, and their
propagational existence includes translational and rotational components},
where the rotational components should, in our view, be connected with the
mutual local energy-momentum exchange between the two subsystems.

Compare to the standard view on electromagnetic field objects the obove
numbered basic properties differ essentially in the final two: standard
relativistic free field electrodynamics formally recognizes the two subsystem
structure, but no recognizable and addmissible changes of each of the two
subsystems, connected with energy-momentum exchange between $F$ and $*F$, is
allowed by the equations $\mathbf{d}F=0, \mathbf{d}*F=0$. In fact, the
divergence of the introduced by Minkowski stress-energy-momentum tensor reads:
$$
Q_\mu^\nu=-\frac12\big[F_{\mu\sigma}F^{\nu\sigma}+
(*)F_{\mu\sigma}(*F)^{\nu\sigma}\big]  \ \rightarrow \ \
\nabla_\nu
Q^\nu_\sigma=\frac12\big{[}F^{\mu\nu}(\mathbf{d}F)_{\mu\nu\sigma}+
(*F)^{\mu\nu}(\mathbf{d}*F)_{\mu\nu\sigma}\big{]}.
$$
Therefore, if the changes of $F$ and $*F$,
represented in classical relativistic electrodynamics by $\mathbf{d}F$ and
$\mathbf{d}*F$, should be recognizable, and so formally to have {\it
tensor nature}, they must NOT be zero in general, but the required equations
$\mathbf{d}F=0, \ \mathbf{d}*F=0$ forbid this. Moreover, if
non-zero admissible changes $\mathbf{d}F$ and $\mathbf{d}*F$ are allowed, they
may be appropriately used in describing the local energy-momentum exchange
between $F$ and $*F$, justifying in this way the intrincic
translational-rotational dynamical structure of a time stable and spatially
propagating electromagnetic field object.

As for the property {\bf 1}, our understanding is that the Minkowcki
energy-momentum tensor $Q$ must be {\it isotropic/null}, i.e.,
$Q_{\mu\nu}Q^{\mu\nu}=0$, and this algebraic equation should be appropriately
justified, respected and used, not just noted. In view of the importance we pay
to it, we give in the next section proof of the Rainich identity, which shows
clearly how it is connected to the invariance properties of free
electromagnetic field objects.

\section{The Rainich identity}
We are going to sketch a proof of the important Rainich identity [6],[7],[8] in
view of its appropriate use in studying the eigen properties of the
electromagnetic energy-momentum tensor on Minkowski space-time
$M=(\mathbb{R}^4,\eta), sign(\eta)=(-,-,-,+)$, and the Hodge $*$ is defined by
$\alpha\wedge *\beta=(-1)^{ind(\eta)}\eta(\alpha,\beta)\omega$, where
$ind(\eta)=3$ and $\omega=dx\wedge dy\wedge dz\wedge d\xi, \ \ \xi=ct$ .

The following relations are
easily verified:
\begin{eqnarray*}
\frac12F_{\alpha\beta}F^{\alpha\beta}id_\mu^\nu&=&F_\mu\,^\sigma
F^\nu\,_\sigma-(*F)_\mu\,^\sigma (*F)^\nu\,_\sigma=
[F\circ F-(*F)\circ (*F)]_\mu^\nu\\
\frac14F_{\alpha\beta}(*F)^{\alpha\beta}id_\mu^\nu&=&
F_\mu\,^\sigma (*F)^\nu\,_\sigma=[F\circ(*F)]_\mu^\nu=
[(*F)\circ F]_\mu^\nu \\
Q_\mu^\nu&=&-\frac12\left[F\circ F+(*F)\circ(*F)\right]_\mu^\nu=
\frac14F_{\alpha\beta}F^{\alpha\beta}id_\mu^\nu-F_\mu\,^\sigma F^\nu\,_\sigma.
\end{eqnarray*}
 Now for the composition $Q\circ Q$ we obtain
\begin{eqnarray*} Q\circ
Q&=&\frac14\Big[F\circ F\circ F\circ F+F\circ F\circ(*F)\circ(*F)\\
&+&(*F)\circ(*F)\circ F\circ F+(*F)\circ(*F)\circ(*F)\circ(*F)\Big]\\
&=&\frac14\Big[F\circ F\circ F\circ F+(*F)\circ(*F)\circ(*F)\circ(*F)+
2(*F)\circ(*F)\circ F\circ F\Big].
\end{eqnarray*}
Making use of the above identities we obtain
\begin{eqnarray*}
F\circ F\circ F\circ
F&=&\frac14(F.F)^2id+\frac{1}{16}(F.*F)^2id+\frac12(F.F)(*F)\circ(*F)\\
(*F)\circ(*F)\circ(*F)\circ(*F)&=&\frac{1}{16}(F.*F)^2id-\frac12(F.F)(*F\circ
*F)\\
2(*F)\circ(*F)\circ F\circ F&=&\frac18(F.*F)^2id,
\end{eqnarray*}
where $(F.F)=F_{\alpha\beta}F^{\alpha\beta}$ and
$(F.*F)=F_{\alpha\beta}(*F)^{\alpha\beta}$. Summing up we get to the Rainich
relation
$$ Q\circ
Q=\frac14\left[\left(\frac12F.F\right)^2+\left(\frac12F.*F\right)^2\right]id=
\frac14\left[I_1^2+I_2^2\right]id .
$$
Clearly, since $tr(id)=4$, we obtain $$ Q_{\mu\nu}Q^{\mu\nu}=I_1^2+I_2^2. $$
Now the eigen relation $Q^\mu_\nu X^\nu=\lambda\,X^\mu$ gives the eigen values
$$
\lambda_{1,2}=\pm\frac12\sqrt{I_1^2+I_2^2}.
$$

We recall now that under the duality transformation
\begin{eqnarray*}
F'&=& F\mathrm{cos}\,\alpha-*F\mathrm{sin}\,\alpha\\
*F'&=& F\mathrm{sin}\,\alpha+*F\mathrm{cos}\,\alpha
\end{eqnarray*}
the energy-momentum tensor stays invariant, but the two invariants $(I_1,I_2)$
keep their values {\it only if they are zero}: $I_1=I_2=0$. Hence, the only
dually invariant eigen direction $\bar{\zeta}$ of the energy-momentum tensor
must satisfy $Q^\mu_\nu\bar{\zeta}^\nu=0$, where $Q$ must satisfy
$det||Q_\mu^\nu||=0$ and $Q\circ Q=0$, i.e. $Q$ becomes {\it boundary map}.
Under these conditions the field $(F,*F)$ is usually called {\it null} field.

We would like specially to note the conformal invariance of the restriction
of the Hodge $*$ to 2-forms. In fact, $\eta'=f^2\eta, f(a)\neq 0, a\in M$,
and $\eta$ generate the same $*$:
$$
*'F=\frac 12 F_{\mu\nu}*'(dx^\mu\wedge dx^\nu)=
-\frac 12 F_{\mu\nu}\eta'^{\mu\sigma}\eta'^{\nu\tau}
\varepsilon_{\sigma\tau\alpha\beta}\sqrt{|det\,\eta'|}
dx^{\alpha}\wedge dx^{\beta}
$$
$$
=-\frac 12 F_{\mu\nu}f^{-4}\eta^{\mu\sigma}\eta^{\nu\tau}
\varepsilon_{\sigma \tau\alpha\beta}f^4 \sqrt{|det\,\eta|}
dx^{\alpha}\wedge dx^{\beta}=*F.
$$
It follows that the stress-energy-momentum tensor $Q_{\mu}^{\nu}$ transforms to
$f^{-4}Q_{\mu}^{\nu}$ under such conformal change of the metric $\eta$.

\section{Some basic properties of null fields}

We begin with recalling that a null local isometry vector field $\bar{\zeta}$
on Minkowski space-time generates null geodesics. In fact, $\bar{\zeta}$
must satisfy $\bar{\zeta}^2=\bar{\zeta}^\nu\zeta_\nu=0$ and
$\nabla_\mu\zeta_\nu+\nabla_\nu\zeta_\mu=0$. From these relations it follows
that
$$
\bar{\zeta}^\mu\nabla_\mu\zeta_\nu+\frac12\nabla_\nu(\bar{\zeta}^\mu\zeta_\mu)=
\bar{\zeta}^\mu\nabla_\mu\zeta_\nu=0.
$$
All free null fields $(F,*F)$, by definition, satisfy
$$
Q_{\mu\nu}Q^{\mu\nu}=0, \ \ \ \text{i.e.}, \ \ \
I_1=\mathbf{B}^2-\mathbf{E}^2=0, \ \ \ I_2=2\mathbf{E}.\mathbf{B}=0,
$$
and {\it in the frame of special relativity the Minkowski metric $\eta$ does not
change in presence of electromagnetic field objects}. Therefore, the null
isometry vector fields and the corresponding geodesics appear as attractive
formal objects to be used in describing the dynamical behavior of the objects
considered. The remarkable two properties of null EM-fields, following from
$Q_{\mu\nu}Q^{\mu\nu}=0$, are that such $Q$ have only zero eigen values and that
they admit unique null eigen direction locally represented by the vector field
$\bar{\zeta}$. As for the eigen vectors of $F$, $*F$ under null $Q$, i.e. when
$I_1=I_2=0$, then all eigen values of $F$ and $*F$ are also equal to zero and
it can be shown [9], that there exists {\it just one common for F, *F, and
Q isotropic eigen direction, defined by the isotropic vector} $\bar{\zeta},
\bar{\zeta}^2=0$, $\bar{\zeta}$ is local isometry for $\eta$, and all other
eigen vectors are space-like.

Thus, the availability of a null electromagnetic field allows to introduce
corresponding sheaf of null geodesics, and this sheaf defines a sheaf of
2-dimensional space-like 2-planes $\mathcal{P}$ orthogonal to $\bar{\zeta}$.
The set of these space-like 2-planes defines 2-dimensional foliation of
Minkowski space-time, and each such 2-plane is integral manifold of geometric
distribution defined by the representatives of the electric and magnetic vector
fields $\mathbf{E},\mathbf{B}$ as tangent to the corresponding 2-plane.

The above considerations allow to choose a global coordinate coframe on $M$ as
follows [9]: $dx$ and $dy$ to determine coframe on each integral 2-dimensional
plane of the distribution, $dz$ to be spatially orthogonal to $dx$ and $dy$,
and $d\xi=c\,dt$ to denote the time coframe 1-form. Denoting further
$\tilde{\eta}(\mathbf{E})\equiv A$, $\tilde{\eta}(\mathbf{B})\equiv A^*$, the
zero values of the two invariants $(I_1,I_2)$ allow to consider the coframe
$(A, A^*, dz,d\xi)$, where
$$
A=u\,dx+p\,dy, \ \
A^*=-\varepsilon\,p\,dx+\varepsilon\,u\,dy , \ \ \varepsilon=\pm 1,
$$
$(u,p)$ are two functions on $M$, as formal image of our field object.
The $\eta$-corresponding image object looks like
$$
\bar{A}=-u\,\frac{\partial}{\partial
x} - p\,\frac{\partial}{\partial y}; \ \
\bar{A^*}=\varepsilon\,p\,\frac{\partial}{\partial x} -
\varepsilon\,u\,\frac{\partial}{\partial y}; \ \
\varepsilon\frac{\partial}{\partial z}; \ \ \frac{\partial}{\partial \xi}\cdot
$$
The eigen null vector field $\bar{\zeta}$ and its $\eta$-coimage $\zeta$ look
like in these coordinates as foloows:
$$
\bar{\zeta}=-\varepsilon\frac{\partial}{\partial z} +
\frac{\partial}{\partial \xi}, \ \ \  \zeta=\varepsilon\,dz+d\xi, \ \
\ \ \varepsilon=\pm 1.
$$
The only non-zero componenets of $Q_{\mu}^{\nu}$ in the induced coordinate frame
$(dx^\mu\otimes\frac{\partial}{\partial x^{\nu}})$ are
$$
Q_{4}^{4}=-Q_3^3=\varepsilon Q_3^4=-\varepsilon Q_4^3=
\frac12(|A|^2+|A^*|^2)=|A|^2=u^2+p^2.
$$
The two 2-forms $F$ and $*F$ look as follows
$$
F=A\wedge\zeta, \ \ \ *F=A^*\wedge\zeta.
$$
Further the coressponding coordinate system will be called $\zeta$-adapted for
short.

We note also the following specific properties of a {\it null} EM-field:

	1. It is determined just by two functions, denoted here by
$u(x,y,z,\xi),p(x,y,z,\xi)$.

	2. It is represented by two algebraically interconnected
through the Hodge $*$-operator and locally recognizable subfields $(F,*F)$
{\it carrying always the SAME stress-energy-momentum}:
$$
I_1=\frac12F_{\mu\nu}F^{\mu\nu}=0
\ \Rightarrow F_{\mu\sigma}F^{\nu\sigma}=(*F)_{\mu\sigma}(*F)^{\nu\sigma}.
 $$

	3. The following relations hold:
$$
i_{\bar{\zeta}}F=i_{\bar{\zeta}}(*F)=0, \
 \ i_{\bar{A}}(*F)=i_{\bar{A}^*}F=0,
$$
where $i_X$ denotes the interior product by the vector $X$.
Hence, $\bar{A}^*$ is eigen
vector of $F$, and $\bar{A}$ is eigen vector of $*F$.

Other two interesting properties of these $F=A\wedge\zeta$ and
$*F=A^*\wedge\zeta$ are the folowing. Consider the $TM$-valued 1-forms
$A\otimes\bar{\zeta}$ and $A^*\otimes\bar{\zeta}$ and compute the corresponding
{\it Fr$\ddot{o}$licher-Nijenhuis} brackets
$[A\otimes\bar{\zeta},A^*\otimes\bar{\zeta}]$  and
$[A\otimes\bar{\zeta},A\otimes\bar{\zeta}]$ (see [10]).
We obtain
$$
[A\otimes\bar{\zeta},A^*\otimes\bar{\zeta}]=
[A^*\otimes\bar{\zeta},A\otimes\bar{\zeta}]
=-\frac12\varepsilon\big[(u^2+p^2)_{\xi}-\varepsilon(u^2+p^2)_z\big]
dx\wedge dy\otimes\bar{\zeta};
$$
$$
[A\otimes\bar{\zeta},A\otimes\bar{\zeta}]=
-[A^*\otimes\bar{\zeta},A^*\otimes\bar{\zeta}]
$$
$$
=[u(p_\xi-\varepsilon\,p_z)-
p(u_\xi-\varepsilon\,u_z)]dx\wedge dy\otimes\bar{\zeta}.
$$

The coressponding Schouten brackets $[\bar{F},\bar{F}]$ and
$[\bar{F},\bar{*F}]$ give
$$
[\bar{F},\bar{F}]=[\bar{A}\wedge\bar{\zeta},\bar{A}\wedge\bar{\zeta}]
$$
$$
=-\varepsilon\big[u(p_{\xi}-\varepsilon p_{z})-p(u_{\xi}-\varepsilon
u_{z})\big]\frac{\partial}{\partial x}\wedge
\frac{\partial}{\partial y}\wedge\frac{\partial}{\partial z}
$$
$$
+\big[u(p_{\xi}-\varepsilon p_{z})-p(u_{\xi}-\varepsilon
u_{z})\big]
\frac{\partial}{\partial x}\wedge
\frac{\partial}{\partial y}\wedge\frac{\partial}{\partial\xi},
$$
\vskip 0.3cm
$$
[\bar{F},\bar{*F}]=[\bar{A}\wedge\bar{\zeta},\bar{A^*}\wedge\bar{\zeta}]
$$
$$
=\frac12\big[(u^2+p^2)_\xi-\varepsilon(u^2+p^2)_z\big]
\frac{\partial}{\partial x}\wedge
\frac{\partial}{\partial y}\wedge\frac{\partial}{\partial z}
$$
$$
-\frac12\varepsilon\big[(u^2+p^2)_\xi-\varepsilon(u^2+p^2)_z\big]
\frac{\partial}{\partial x}\wedge
\frac{\partial}{\partial y}\wedge\frac{\partial}{\partial\xi}.
$$

\section{The new equations and their properties}
\subsection{Mathematical identification of the field}

As we mentioned, the physical object we are going to mathematically describe by
means of a {\it system of partial differential equations} on Minkowski
space-time satisfies the condition: its physical appearance is formally
represented by $(F,*F)$, its dynamical appearance is formally represented by
the energy-momentum tensor $Q$, which satisfies the relations:
$Q_{\mu\nu}Q^{\mu\nu}=I_1^2+I_2^2=0 \rightarrow
F_{\mu\sigma}F^{\nu\sigma}=(*F)_{\mu\sigma}(*F)^{\nu\sigma},
\nabla_{\nu}Q_\mu^\nu=0$, hence, it allows to be viewed as built of
two recognizable subsystems which carry the same local stress-energy-momentum.
It follows from this view that, if these two subsystems interact, i.e.,
exchange energy-momentum, they {\it must be in a permanent local dynamical
equilibrium: making use of their $\eta$ co-images they permanently and directly
exchange energy-momentum in equal quantities without available local
interaction energy}. \vskip 0.2cm The apperant forms of the two space-time
recognizable subsystems in our $\zeta$-adapted coordinate system allow they to
be mathematicaly identified by two subdistributions in the tangent bundle of
Minkowski space-time and to make use of their $\eta$-codistributions, so that,
{\it no admissible coordinate/frame change to result in nullifying locally or
globaly of any of these two mathematical images of the two recognizable
subsystems}.

It deserves also noting that the two subsystems recognize each other in two
ways: algebraically - through the Hodge $*$-operator; and differentially -
through the allowed local energy-momentum exchange.

Of course, these two kinds of contact between the two mathematical
representatives should be {\it physically motivated}, i.e. they should reflect
some physical appearances of the field object carrying such dynamical structure.

We give some general preview consideration.

From algebraic point of view the exterior powers of a vector space naturally
separate lineary independent elements: $x\wedge y$ is not zero only if $x\neq
\lambda y$. So, if our physical object of interest is built of $p$ interacting
and recognizable time-stable subsystems, it seems natural to turn to the
exterior algebras built over corresponding couple of dual linear spaces. This
view allows the well known concepts of interior products between $p$-vectors
and $q$-forms [11,12] to be correspondingly respected and physically interpreted
as flows, in other words, as quantitative measures of energy-momentum exchange.

It deserves also noting that any choice of {\it decomposable}
$p$-vector $\Phi$ over a linear space $E^n$ automatically
 defines a $p$-dimensional subspace $E^p_{\Phi}\subset
E^n$. Now, making use of the Poincare isomorphism $D^p: E^p\rightarrow
E_{n-p}^*$ [11] we can determine the object $D^p(\Phi)$, which defines a
$(n-p)$-dimensional subspace $D^p(E^{p}_{\Phi})\subset (E^n)^*$, where
$(E^n)^*$ is the dual for $E^n$ space. Two more subspaces, namely,
$(E^p_{\Phi})^*\subset (E^n)^*$, which is the dual to $E^p_{\Phi}$, and
$(D^p(E^{p}_{\Phi}))^*\subset E^n$, which is the dual to $D^p(E^{p}_{\Phi})$,
immediately appear.

The above pure algebraic facts may be carried to tangent/cotangent bundles
of a manifold $M^n$ through the well known concept of {\it distribution}
[10,13,14], i.e., a sub-bundle of a tangent bundle.
A basic tradition in physics, however, is measuring distance, which requires
metrics/pseudometrics $g$ in theory. This allows to make use of the
corresponding $g$-defined isomorphisms in the tensor algebra on a manifold, in
particular, in the corresponding exterior subalgebras, composed with the
Hodge-$*_{g}$ operator, as an appropriate substitutes of Poincare isomorphisms
when possible. So, if the metric is known somehow, this leads to appropriate
for theoretical physics explicit connection between tensors and co-tensors, in
particular, between $p$-vector fields and $p/(n-p)$-differential forms. Now, if
we have come to the conclusion that a physical system $\mathcal{A}$ may be
represented by appropriate distribution $\Delta$  on $M^n$, we should keep in
mind that the same $\Delta$ may be defined by various appropriate non-singular
multivector fields. This moment is very important when we introduce symmetries
of $\Delta$ [14] through Lie derivative: a vector field $X$ is a symmetry of
$\Delta$ if for every vector section $Y$ of $\Delta$ we have that $[X,Y]=L_XY$
lives in $\Delta$. This suggests to understand the space-time evolution of
$\mathcal{A}$ along appropriate time-like or isotropic/null vector fields in
the case of Minkowski-like metrics.

This concerns also the case when we try to look inside $\mathcal{A}$ in order to
distinguish time-stable recognizable subsystems
$\mathcal{A}_1,\mathcal{A}_2,...$ of $\mathcal{A}$, trying to represent them by
corresponding subdistributions $\Delta_1,\Delta_2,...$ of $\Delta$. Now, the
time stability of $\mathcal{A}$ should, at least partially, depend on the available
local interactions among the subsystems of $\mathcal{A}$, which interactions
could be represented in terms of the curvature forms $\Omega_i$ of the
corresponding nonintegrable subdistributions $\Delta_i$, provided the
values of $\Omega_i$ are inside $\Delta$.

We briefly give now the formal picture.

Recall that the duality between the two $n$-dimensional vector spaces $E$ and
$E^*$ allows to distinguish the following antiderivation [11]. Let $h\in E$,
then we obtain the antiderivation $i(h)$, or $i_h$, in $\Lambda(E^*)$ of degree
$(-1)$ (sometimes called substitution/contraction operator, interior product,
insertion operator)
 according to:
$$
i(h)(x^{*1}\wedge\dots\wedge
x^{*p})=\sum_{i=1}^{p}(-1)^{(i-1)}\langle x^{*i},h\rangle
x^{*1}\wedge\dots\wedge\hat{x^{*i}}\wedge
\dots\wedge x^{*p}.
$$
Clearly, if $u^*\in \Lambda^p(E^*)$ and $v^*\in\Lambda(E^*)$ then
$$
i(h)(u^*\wedge v^*)=(i(h)u^*)\wedge v^*+(-1)^pu^*\wedge i(h)v^*.
$$
Also, we get
$$
i(h)u^*(x_1,\dots,x_{p-1})=u^*(h,x_1,\dots,x_{p-1}), \
$$
$$
i(x)\circ i(y)=-i(y)\circ i(x).
$$

This antiderivation is extended [11,12] to a mapping $i(h_1\wedge\dots\wedge
h_p): \Lambda^m(E^*)\rightarrow\Lambda^{(m-p)}(E^*)$, $m\geqq p$, according to
$$
i(h_1\wedge h_2\wedge\dots\wedge h_p)u^*=i(h_p)\circ\dots\circ i(h_1)\,u^*.
$$
Note that this extended mapping is not an antiderivation except for $p=1$.

This mapping is extended by linearity to multivectors and exterior forms which
are linear combinations.

Let now the (multi)vector field $T_1$ live in a distribution $\Delta^p(M)$ on a
manifold $M^n, p<n$, with corresponding codistribution $\Delta^p_{*}(M)$, and
the (multi)vector field $T_2$ live in a distribution $\Delta^q(M), q<n$, with
corresponding codistribution $\Delta^q_{*}(M)$. If the flow
$i_{T_1}\Delta^q_{*}(M)$ of $T_1$ across $\Delta^q_{*}(M)$ is not zero, i.e.,
the flows of $T_1$ across some $q$-forms in $\Delta^q_{*}(M)$ are NOT zero,
and the flow $i_{T_2}\Delta^p_{*}(M)$ of $T_2$ across $\Delta^p_{*}(M)$ is not
zero, i.e., the flows of $T_2$ across some $p$-forms in $\Delta^p_{*}(M)$ are
NOT zero, then these two distributions may be called {\it interacting
partners}. Clearly, this situation can be understood in terms of curvature
forms if the (multi)vectors are composed of Lie brackets of
vector fields living correspondingly in $\Delta^p(M)$, or $\Delta^q(M)$. For
example, if $X,Y$ live in $\Delta^p(M)$ and $\alpha$ lives in
$\Delta^{(q=1)}_{*}(M)$, and $\Delta^p(M)$ and $\Delta^q(M)$ are not
intersecting: $\langle\Delta^q_{*}(M),\Delta^p(M)\rangle=0$, then the flow of
the 2-vector $X\wedge Y$ across $\mathbf{d}\alpha$ reduces to the flow of
$[X,Y]$ across $\alpha$ since $\mathbf{d}\alpha(X,Y)=X\langle\alpha,Y\rangle-
Y\langle\alpha,X\rangle-\alpha([X,Y])=-\alpha([X,Y])$, and
$\langle\alpha,X\rangle=\langle\alpha,Y\rangle=0$.

The Lie derivative of a differential $q$-form  $\alpha$ along a $p$-multivector
$T$ is naturally defined by [12]
$$
L_{T}\alpha=\mathbf{d}\langle i_T\alpha\rangle-
(-1)^{deg T.deg \mathbf{d}}i_T\mathbf{d}\alpha, \ \ \ deg(\mathbf{d}=1.
$$

We construct now the $\varphi$-extended insertion operator $i^{\varphi}_T$
on $M$. Let $(E_1,E_2,F)$ be three linear spaces, $\{e_i\}$ and $\{k_j\}$ be
bases of $E_1$ and $E_2$ respectively, $T=\mathfrak{t}^i\otimes
e_i\in\mathfrak{X}^q(M,E_1)$ be a non-decomposable $E_1$-valued q-vector,
$\mathfrak{t}^i$ represent corresponding distributions with dual
codistributions $\mathfrak{t^*}^i$, $\Phi=\alpha^j\otimes
k_j\in\Lambda^p(M,E_2)$ be a non-decomposable $E_2$-valued p-form, $\alpha^j$
represent corresponding codistributions with $\alpha^j_{*}$ their dual
distributions, with $q\leq p$, and $\varphi:E_1\times E_2\rightarrow F$ be a
bilinear map. Now we define the {\it $\varphi$-algebraic flow} of $T$ across
$\Phi$: $i^{\varphi}_T\Phi\in\Lambda^{p-q}(M,F)$ as follows:
$$
i^{\varphi}_T\Phi= i_{\mathfrak{t}^i}\alpha^j\otimes\varphi(e_i,k_j), \ \ \
i=1,2,...,dim(E_1), \ j=1,2,...,dim(E_2).
$$
Hence, we can define the $\varphi$-extended Lie derivative of $T$ across $\Phi$:
$$
\mathcal{L}^{\varphi}_T:
\Lambda^p(M,E_2)\times\mathfrak{X}^q(M,E_1)\rightarrow\Lambda^{p-q+1}(M,F)
$$
as follows
$$
\mathcal{L}^{\varphi}_T(\Phi)=
\mathbf{d}\circ i^{\varphi}_T\Phi-
(-1)^{deg(T).deg(\mathbf{d})}i^{\varphi}_T\circ\mathbf{d}\Phi.
 $$
Accordingly, $T$ will be called {\it algebraic} $\varphi$-symmetry of $\Phi$ if
$i^{\varphi}_T\Phi$ is a {\it constant} element of $\Lambda^{p-q}(M,F)$, and
{\it differential} $\varphi$-symmetry of $\Phi$, if
$\mathcal{L}^{\varphi}_T(\Phi)=0$. Also, $(T,\Phi)$ may be called
{\it in mutual local contact} if there are at least two differential flows
$\mathcal{L}^{\varphi}_{\mathfrak{t^i}}\alpha^{j}$ and
$\mathcal{L}^{\varphi}_{\alpha^j_{*}}\mathfrak{t}^{i*}$
which are NOT zero. Clearly, the
bilinear map $\varphi$ is meant to distinguish those couples of distributions
which are in contact with each other through their curvature forms [15].

Let's see now what Minkowski space-time manifold $M=(\mathbb{R}^4,\eta)$ may
offer in this direction.

The basic mathematical object on $M$ is its pseudometric tensor $\eta$, it
defines the mathematical procedure that corresponds to the experimental
procedure for measuring space distance making use of light signals. In terms of
$\eta$ we algebraically define 4-volume on $M$ and appropriate linear
isomorphisms in the tensor algebra over $M$. Also, the exterior algebra of
differential forms can be equiped with the $\eta$-defined linear isomorphism
between $\Lambda^p(M)$ and $\Lambda^{4-p}(M)$ by the Hodge $*_p$-operator. In
view of the existence of $\eta$, we are going to make use of the
Hodge-$*_{\eta}$ and of the $\tilde{\eta}$-isomorphisms which will serve as good
substitutes for the Poincare isomorphisms $D^p$.

These remarks clearly support the opinion that Minkowski has made appropriate
steps towards mathematical identification of the two physical subsystems of an
electromagnetic free object. Also, the above remarks suggest to slightly modify
the Minkowski choice for mathematical identification of the field as follows:
the two differential forms $(F,*F)$ to be unified as one $\mathbb{R}^2$-valued
differential 2-form $\Omega$ on $M$, and their $\eta$-images to be unified as
one $\mathbb{R}^2$-valued $2$-vector field $\bar{\Omega}$:
$$
\Omega=F\otimes e_1+*F\otimes e_2, \ \ \
\bar{\Omega}=\bar{F}\otimes e_1+\bar{*F}\otimes e_2 ,
$$
where $(e_1,e_2)$ is a basis of the vector space
$\mathbb{R}^2$, and the bar over $F$ and $*F$ denotes the coressponding
$\tilde{\eta}$-images. In this way, recalling the divergence expression for
$Q_\mu^\nu$, we may directly turn our attention to the following four
differential mutual flows
$$
i_{\bar{F}}\mathbf{d}F, \ \ \
i_{\bar{*F}}\mathbf{d}*F , \ \ \ i_{\bar{F}}\mathbf{d}*F, \ \ \
i_{\bar{*F}}\mathbf{d}F
$$
as appropriate {\it local} energy-momentum balance quantities.

\subsection{Dynamical equations}

According to the above assumptions an EM-{\it null} field object
must survive through space-time propagation during which it has to keep its
structure through establishing and supporting internal dynamical equilibrium
between its two recognizable subsystems. Our mathematical interpretation of
this vision differs substantially from that of Maxwell-Minkowski, simply
speaking, it consists in considering $\bar{\Omega}$ as a $\vee$-extended {\it
algebraic} and {\it differential} symmetry of $\Omega$, where "$\vee$" denotes
here symmetrised tensor product applied to the vector space $\mathbb{R}^2$:
$$
i^{\vee}_{\bar{\Omega}}\Omega
=\mathfrak{C}\in\Lambda_{const}^{p-q}(M,\mathbb{R}^2),
\ \ \ \mathcal{L}^{\vee}_{\bar{\Omega}}\Omega=0.
$$

Explicitly, the differential $\vee$-symmetry gives:
\begin{eqnarray*}
\mathcal{L}_{\bar{\Omega}}^{\vee}\Omega&=& \left[\mathbf{d}\langle
F,\bar{F}\rangle-i_{\bar{F}}(\mathbf{d}F) \right]\otimes e_1\vee e_1\\
&+&\left[\mathbf{d}\langle
*F,\bar{*F}\rangle-i_{\bar{*F}}(\mathbf{d}*F)\right]\otimes
e_2\vee e_2\\ &+&
\{2\mathbf{d}\langle F,\bar{*F}\rangle-[i_{\bar{*F}}(\mathbf{d}F)+
i_{\bar{F}}(\mathbf{d}*F)]\}\otimes e_1\vee e_2=0 .
\end{eqnarray*}
\vskip 0.2cm
{\bf Remark}. We have chosen the $"\vee"$-extension of the Lie derivative
paying due respect to the entire symmetry between the two components $F$ and
$*F$ and to the dynamical inter-equilibrium they keep during propagation.
\vskip 0.2cm
The equations we obtain are
\begin{eqnarray*}
&&\mathbf{d}\langle F,\bar{F}\rangle-i_{\bar{F}}(\mathbf{d}F)=0 ,
\\
&&\mathbf{d}\langle
*F,\bar{*F}\rangle-i_{\bar{*F}}(\mathbf{d}*F)=0 ,
\\
&&2\mathbf{d}\langle F,\bar{*F}\rangle-[i_{\bar{*F}}(\mathbf{d}F)+
i_{\bar{F}}(\mathbf{d}*F)]=0.
\end{eqnarray*}

Since in our case the formal identity
$\langle F,\bar{F}\rangle=-\langle *F,\bar{*F}\rangle$ always holds,
summing up the first two equations we obtain
$$
i_{\bar{F}}\mathbf{d}F+i_{\bar{*F}}\mathbf{d}*F=0, \ \ \text{i.e.} \ \ \
F^{\alpha\beta}(\mathbf{d}F)_{\alpha\beta\mu}+
(*F)^{\alpha\beta}(\mathbf{d}*F)_{\alpha\beta\mu}=0, \ \ \ \alpha<\beta,
$$
which coincides with the zero divergence of the standard
and well trusted electromagnetic stress-energy-momentum tensor $Q_\mu^\nu$:
$$
\nabla_\nu Q^\nu_\mu
=\nabla_\nu\Big[-\frac12\big(F_{\mu\sigma}F^{\nu\sigma}+
(*F)_{\mu\sigma}(*F)^{\nu\sigma}\big)\Big]=
F^{\alpha\beta}(\mathbf{d}F)_{\alpha\beta\mu}+
(*F)^{\alpha\beta}(\mathbf{d}*F)_{\alpha\beta\mu}=0, \ \ \alpha<\beta.
$$

From the explicit expression of $Q_\mu^\nu$ in terms of $F$ and $*F$
it is clearly seen that the full stress-energy-momentum of the field is the sum
of the stress-energy-momentum carried by $F$, i.e.
$\frac12F_{\mu\sigma}F^{\nu\sigma}$, and the stress-energy-momentum carried by
$(*F)$, i.e. $\frac12(*F)_{\mu\sigma}(*F)^{\nu\sigma}$.
Now, the algebraic $\vee$-symmetry equation
$$i^{\vee}_{\bar{\Omega}}\Omega=
\frac12F_{\mu\sigma}F^{\mu\sigma}e_1\vee e_1+
\frac12(*F)_{\mu\sigma}(*F)^{\mu\sigma}e_2\vee e_2+
\frac12F_{\mu\sigma}(*F)^{\mu\sigma}e_1\vee e_2=\mathfrak{C}
$$
requires
$$
I_1=\frac12F_{\mu\sigma}F^{\mu\sigma}=
-\frac12(*F)_{\mu\sigma}(*F)^{\mu\sigma}
=\frac12\langle F,\bar{F}\rangle=const,
$$
$$
I_2=\frac12F_{\mu\sigma}(*F)^{\mu\sigma}=\frac12\langle F,\bar{*F}\rangle=
const.
$$

It follows $\mathbf{d}\langle F,\bar{F}\rangle=\mathbf{d}\langle
*F,\bar{F}\rangle=0$, hence, the above equations
$i^{\vee}_{\bar{\Omega}}\Omega=\mathfrak{C}, \ \
\mathcal{L}^{\vee}_{\bar{\Omega}}\Omega=0$, i.e. the algebraic and differential
$\vee$-symmetry of $\Omega$ with respect to $\bar{\Omega}$, give :
\begin{eqnarray*}
&&F^{\alpha\beta}(\mathbf{d}F)_{\alpha\beta\mu}=0, \\
&&(*F)^{\alpha\beta}(\mathbf{d}*F)_{\alpha\beta\mu}=0, \\
&&(*F)^{\alpha\beta}(\mathbf{d}F)_{\alpha\beta\mu}+
F^{\alpha\beta}(\mathbf{d}*F)_{\alpha\beta\mu}=0 , \ \ \alpha<\beta.
\end{eqnarray*}
It is seen that the two equations $\mathcal{L}^{\vee}_{\bar{\Omega}}\Omega=0$
and $i^{\vee}_{\bar{\Omega}}\Omega=\mathfrak{C}$ reduce to
$i^{\vee}_{\bar{\Omega}}\mathbf{d}\Omega=0$.
In this way a permanent local dynamical equilibrium between
the two subsystems formally represented by $F$ and $*F$ is established.

The conformal invariance if these equations follows from the conformal
invariance of $*_{2}$, and every solution $(F,*F)$ realizes the idea for
{\it local dynamical equilibrium}. In terms of the coderivative
$\delta=*\,\mathbf{d}\,*$ we get
\begin{eqnarray*}
&&F_{\mu\nu}(\delta
F)^\nu=0,\\ &&(*F)_{\mu\nu}(\delta *F)^\nu=0,\\ &&(*F)_{\mu\nu}(\delta
F)^\nu+F_{\mu\nu}(\delta *F)^\nu=0.
\end{eqnarray*}
The coordinate-free form of these equations reads:
\begin{eqnarray*}
(*F)\wedge*\mathbf{d}*F &\equiv&\delta
F\wedge *F=0,\\ F\wedge *\mathbf{d}F &\equiv& -\delta *F\wedge F=0,\\ F\wedge
*\mathbf{d}*F+(*F)\wedge*\mathbf{d}F &\equiv&\delta F\wedge F-\delta *F\wedge
*F=0.
\end{eqnarray*}

\section{Some properties of the nonlinear solutions}
Clearly, among the solutions of our equations there are nonlinear ones,
satisfying $\mathbf{d}F\neq 0,
\mathbf{d}*F\neq 0$, or, equivalently,
$\delta *F\neq 0, \delta F\neq 0$.

Further we concentrate on the nonlinear ones.

First we prove that all nonlinear solutions are {\it null}, i.e.,
$\mathfrak{C}=0$.

Recall the relations satisfied by any 2-form $A$ on Minkowski space-time [11]:
$$
A\wedge A=\sqrt{det(A_{\mu\nu})}\,dx\wedge dy\wedge dz\wedge d\xi
, \ \ A\wedge A=-A\wedge *(*A)=\frac12A_{\mu\nu}(*A)^{\mu\nu}dx\wedge dy\wedge
dz\wedge d\xi.
$$
Since all nonlinear solutions $(F,*F)$ satisfy $\delta F\neq
0, \delta *F\neq 0$, from $F_{\mu\nu}(\delta F)^\nu=0, (*F)_{\mu\nu}(\delta
*F)^\nu=0$ it follows that they must satsfy
$$
det||F_{\mu\nu}||=0, \ \ \ det||(*F)_{\mu\nu}||=0, \ \ \text{i.e.}, \ \
I_2=\frac12F_{\mu\nu}(*F)^{\mu\nu}=0.
$$
Further, summing up the three systems of equations, we obtain
$$
(F+*F)_{\mu\nu}(\delta F+\delta *F)^\nu=0.
$$

If now $(\delta F+\delta *F)^\nu\neq 0$, then
$$
0=\mathrm{det}||(F+*F)_{\mu\nu}||=
\left [-F_{\mu\nu}F^{\mu\nu}\right ]^2=(2I_1)^2.
$$

If $\delta F^\nu=-(\delta *F)^\nu\neq 0$, we sum up the first two systems and
obtain $(*F-F)_{\mu\nu}(\delta *F)^\nu=0$. Consequently,
$$
0=\mathrm{det}||(*F-F)_{\mu\nu}||=
\left [F_{\mu\nu}F^{\mu\nu}\right ]^2=(2I_1)^2.
$$
This completes the proof. Hence,
the corresponding $\mathfrak{C}$ is zero: $\mathfrak{C}=0$ .

Now, according to the Rainich identity, this is equivalent to
$Q_{\mu\nu}Q^{\mu\nu}=0$.
In view of this, from the formal identity
$$
I_1=\frac12F_{\alpha\beta}F^{\alpha\beta}\delta^\nu_\mu=
F_{\mu\sigma}F^{\nu\sigma}-(*F)_{\mu\sigma}(*F)^{\nu\sigma}
$$
it follows that the two subsystems carry the same stress-energy-momentum:
$F_{\mu\sigma}F^{\nu\sigma}=(*F)_{\mu\sigma}(*F)^{\nu\sigma}$.

Hence, the above considered
properties of null fields hold for all nonlinear solutions and the
representation $F=A\wedge\zeta\neq 0, *F=A^*\wedge\zeta\neq 0$ is allowed.

This representation says that with every nonlinear solution $(F,*F)$
three geometric 2-dimensional distributions on Minkowski space may be
introduced, such that:

	- the distribution $(\bar{A},\bar{A^*})$ is completely integrable:
$[\bar{A},\bar{A}^*]\wedge\bar{A}\wedge\bar{A}^*=0$.

 - the other two distributions $(\bar{A},\bar{\zeta})$ and
$(\bar{A^*},\bar{\zeta})$ are not completely integrable. The corresponding
nonintegrability relations in terms of the codistributions read
$$
 \mathbf{d}A\wedge A\wedge \zeta= \varepsilon\big[u(p_\xi-\varepsilon
p_z)- p\,(u_\xi-\varepsilon u_z)\big]\omega=\varepsilon\mathbf{R}\,\omega;
$$
$$
\mathbf{d}A^*\wedge A^*\wedge \zeta=                                  
\varepsilon\big[u(p_\xi-\varepsilon p_z)- p\,(u_\xi-\varepsilon
u_z)\big]\omega=\varepsilon\mathbf{R}\,\omega,
$$
where $\omega=dx\wedge dy\wedge dz\wedge d\xi$, and
$\mathbf{R}=\big[u(p_\xi-\varepsilon p_z)- p\,(u_\xi-\varepsilon u_z)]$ denotes
the corresponding curvature. The two curvature forms read
$$
\mathfrak{R}=-\mathbf{d}A^*\otimes\frac{\bar{A^*}}{|\bar{A^*}|^2}
,\ \ \
\mathfrak{R}^*=-\mathbf{d}A\otimes\frac{\bar{A}}{|\bar{A}|^2}.
$$
On the other hand we obtain
$$
\langle
A,[\bar{A},\bar{\zeta}]\rangle=                          
\langle
A^*,[\bar{A^*},\bar{\zeta}]\rangle= \frac12\big[(u^2+p^2)_\xi-\varepsilon
(u^2+p^2)_z\big].
$$
Clearly, the last relation may be put in
terms of the Lie derivative $L_{\bar{\zeta}}$ as
$$
\frac 12L_{\bar{\zeta}}(u^2+p^2)=
-\frac12L_{\bar{\zeta}}\langle A,\bar{A}\rangle=
-\langle A,L_{\bar{\zeta}}\bar{A}\rangle=
-\langle A^*,L_{\bar{\zeta}}\bar{A^*}\rangle.
$$

{\bf Remark}. Further we shall denote $\sqrt{u^2+p^2}\equiv
\phi$.

We notice now that there is a function $\psi(u,p)$ such, that

$$L_{\bar{\zeta}}\psi=
\frac{u(p_\xi-\varepsilon p_z)-p(u_\xi-\varepsilon u_z)}{\phi^2}=
\frac{\mathbf{R}}{\phi^2} .
$$
\vskip 0.4cm

It is immediately verified that $\psi=\arctan\frac pu$ is such one.
\vskip 0.4cm
We note that the function $\psi$ has a natural
interpretation of {\it phase} because of the easily verified now relations
$u=\phi\cos\psi$, $p=\phi \sin\psi $,
and $\phi^2$ acquires the status of energy density. Since
the transformation $(u,p)\rightarrow (\phi,\psi)$ is non-degenerate this allows
to work with the two functions $(\phi,\psi)$ instead of $(u,p)$, and the
equations reduce to $L_{\bar{\zeta}}\phi=0$,
so $\phi(x,y,\xi+\varepsilon z), \varepsilon=\pm1$ and $\psi$ is arbitrary.

From the above we have
$$
\mathbf{R}=\phi^2L_{\bar{\zeta}}\psi= \
\phi^2(\psi_\xi-\varepsilon\psi_z) \ \ \
\rightarrow \ \ L_{\bar{\zeta}}\psi=
\frac{\mathbf{R}}{T(\partial_{\xi},\partial_{\xi})}=
\frac{*\varepsilon(\mathbf{d}A\wedge A\wedge\zeta)}{T(\partial_{\xi},\partial_{\xi})},
$$
where $T(\partial_{\xi},\partial_{\xi})$ is the coordinate-free definition of
the energy density.

 This last formula shows something very important: at
any $\phi\neq 0$ the curvature $\mathbf{R}$ will NOT be zero only if
$L_{\bar{\zeta}}\psi\neq 0$, which admits in principle availability of
rotation. In fact, lack of rotation would mean that $\phi$ and $\psi$ are
running waves along $\bar{\zeta}$. The relation $L_{\bar{\zeta}}\psi\neq 0$
means, however, that rotational properties are possible in general, and some of
these properties are carried by the phase $\psi$. It follows that in such a
case the translational component of propagation along $\bar{\zeta}$ (which is
supposed to be available) must be determined essentially, and most probably,
entirely, by $\phi$.  In particular, we could expect the relation
$L_{\bar{\zeta}}\phi=0$ to hold, and if this happens, then the rotational
component of propagation will be represented entirely by the phase $\psi$, and,
more specially, by the curvature factor $\mathbf{R}\neq 0$, so, the objects we
are going to describe may have compatible translational-rotational dynamical
structure. Finally, this relation may be considered as a definition for the
phase function $\psi$.

Another interesting relations are the following:
$$
\mathfrak{R}(\bar{\zeta},\bar{A})=-\frac{\varepsilon\mathbf{R}}{\phi^2}\bar{A^*},
\ \ \ \mathfrak{R}^*(\bar{\zeta},\bar{A}^*)=
\frac{\varepsilon\mathbf{R}}{\phi^2}\bar{A},
$$
$$
\ \ \ i(\mathfrak{R}(\bar{\zeta},\bar{A}))*F=-
i(\mathfrak{R}^*(\bar{\zeta},\bar{A}^*))F=
i(\bar{*F})\mathbf{d}F=-i(\bar{F})\mathbf{d}*F=\varepsilon\mathbf{R}\zeta,
$$
also, $\delta\,*F\wedge *F=\delta\,F\wedge F=\varepsilon\mathbf{R}*\zeta$.

Hence, the two mutual flows are not zero only if the curvature $\mathbf{R}$ is
not zero, so, we can try the 3-form $\delta\,F\wedge F$ as possible carrier of
 spin properties of the corresponding solution. In fact, it turns out
that the equations require $\phi$ to be running wave along the $z$-coordinate
(in the coordinates used), and arbitrary with respect to the spatial
coordinates $(x,y,z)$:  $\phi(x,y,\xi+\varepsilon\,z)$, so, it may be chosen
spatially finite, and since it defines the energy density of the solution,
then, since $\frac{\partial}{\partial\xi}$ is local isometry, then
 the reduced to the 3\,-space 3\,-form $*(Q_{\mu 4}dx^{\nu})$
may give finite integral energy. Moreover, the choice
$\psi(z)=cos(\kappa\frac{z}{\mathcal{L}_o}+const), \kappa=\pm 1$,
$\mathcal{L}_o=const$ has dimension of length, is also allowed by the
equations, which leads to $\mathbf{d}(\delta\,F\wedge F)=0$. Introducing now
the 3-form
$$
\beta=2\pi\frac{\mathcal{L}^2_o}{c}F\wedge \delta F,
$$
its
reduction to $\mathbb{R}^3$ is $$ \beta_{\mathbb{R}^3}=\kappa
2\pi\frac{\mathcal{L}^2_o}{c} \phi^2 dx\wedge dy\wedge dz. $$

On the two figures below are given two
theoretical examples with $\kappa=-1$ and $\kappa=1$ respectively, at a fixed
moment $t$. For $t\in (-\infty,+\infty)$, the amplitude
function $\phi$ fills in a smoothed out tube around a circular helix of height
$2\pi\mathcal{L}_o$ and pitch $\mathcal{L}_o$, and phase
function $\varphi=\mathrm{cos}(\kappa z/\mathcal{L}_o)$. The solutions
propagate left-to-right along the euclidean coordinate $z$.

\begin{center}
\begin{figure}[ht!]
\centerline{
{\mbox{\psfig{figure=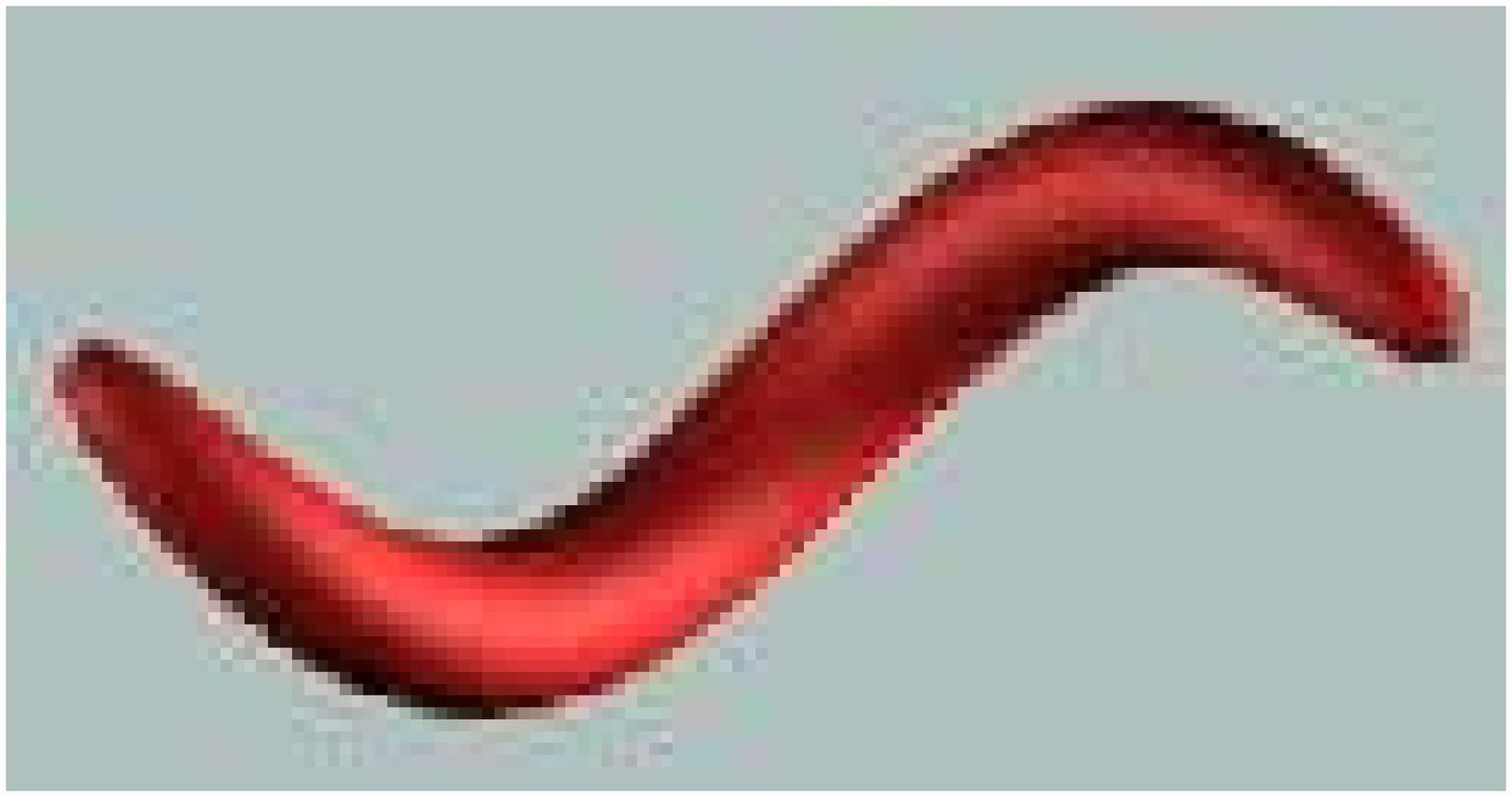,height=1.8cm,width=3.5cm}}
\mbox{\psfig{figure=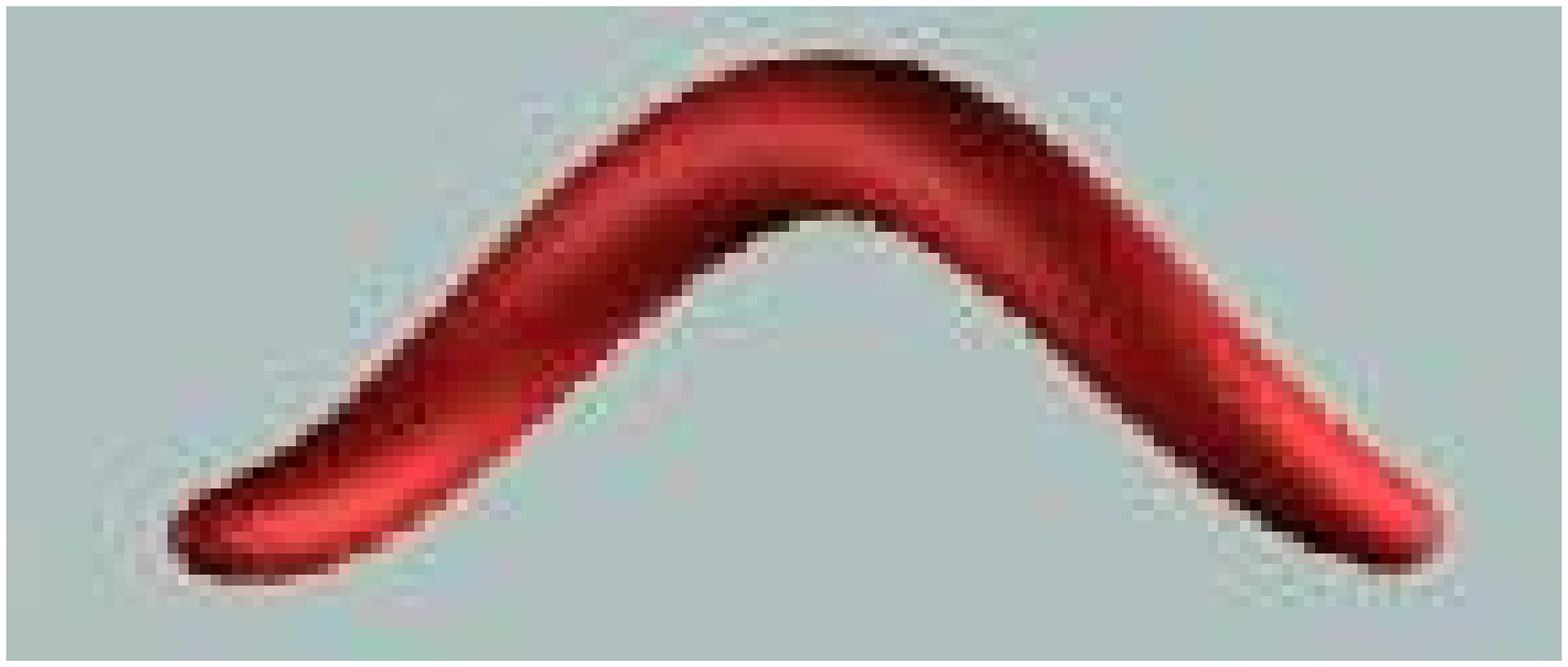,height=1.8cm,width=4.2cm}}
\mbox{\psfig{figure=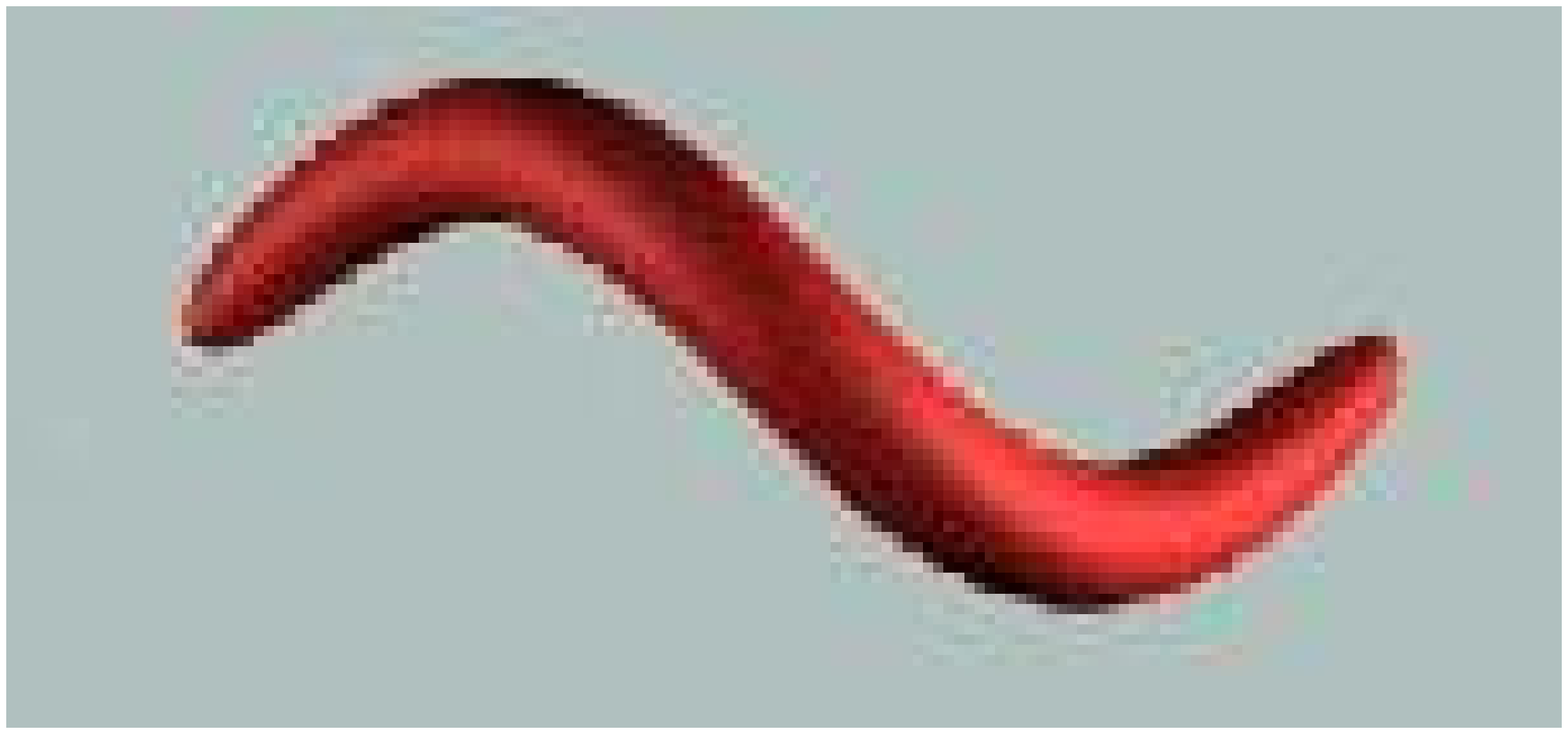,height=1.8cm,width=4.2cm}}}}
\caption{Theoretical example with $\kappa=-1$. The Poynting vector is
directed left-to-right.}
\end{figure}
\end{center}
\begin{center}
\begin{figure}[ht!]
\centerline{
{\mbox{\psfig{figure=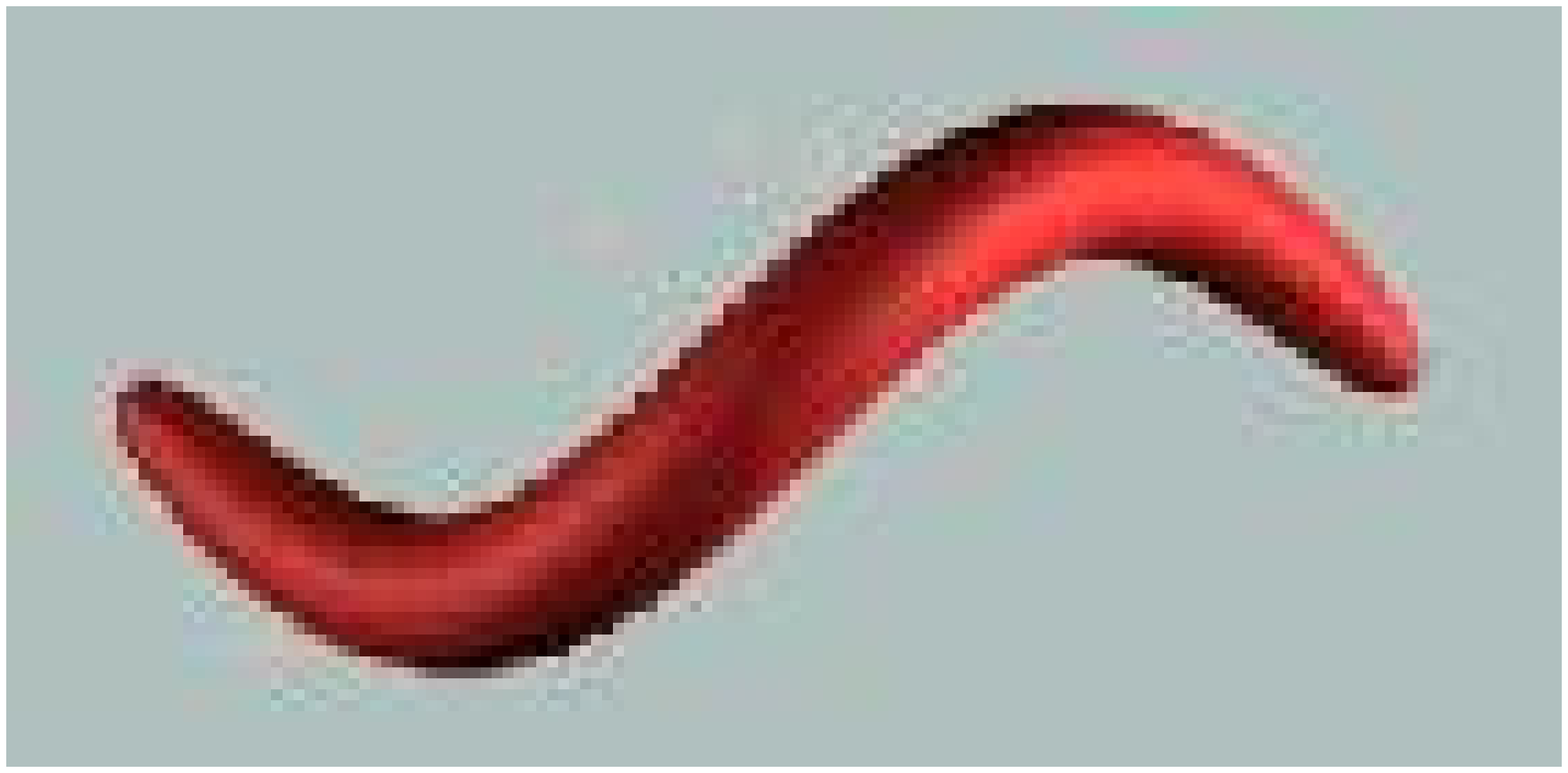,height=1.8cm,width=3.5cm}}
\mbox{\psfig{figure=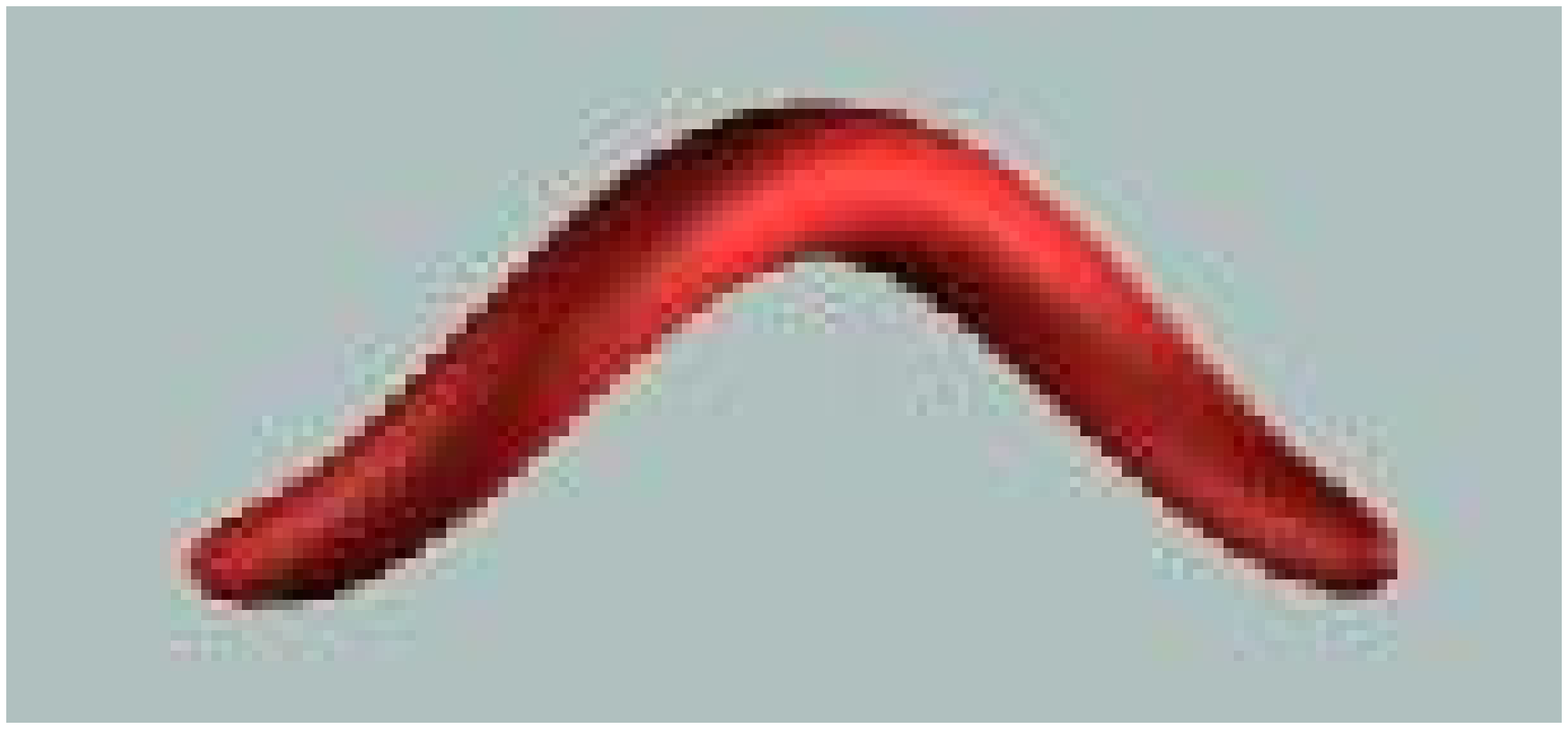,height=1.8cm,width=4.2cm}}
\mbox{\psfig{figure=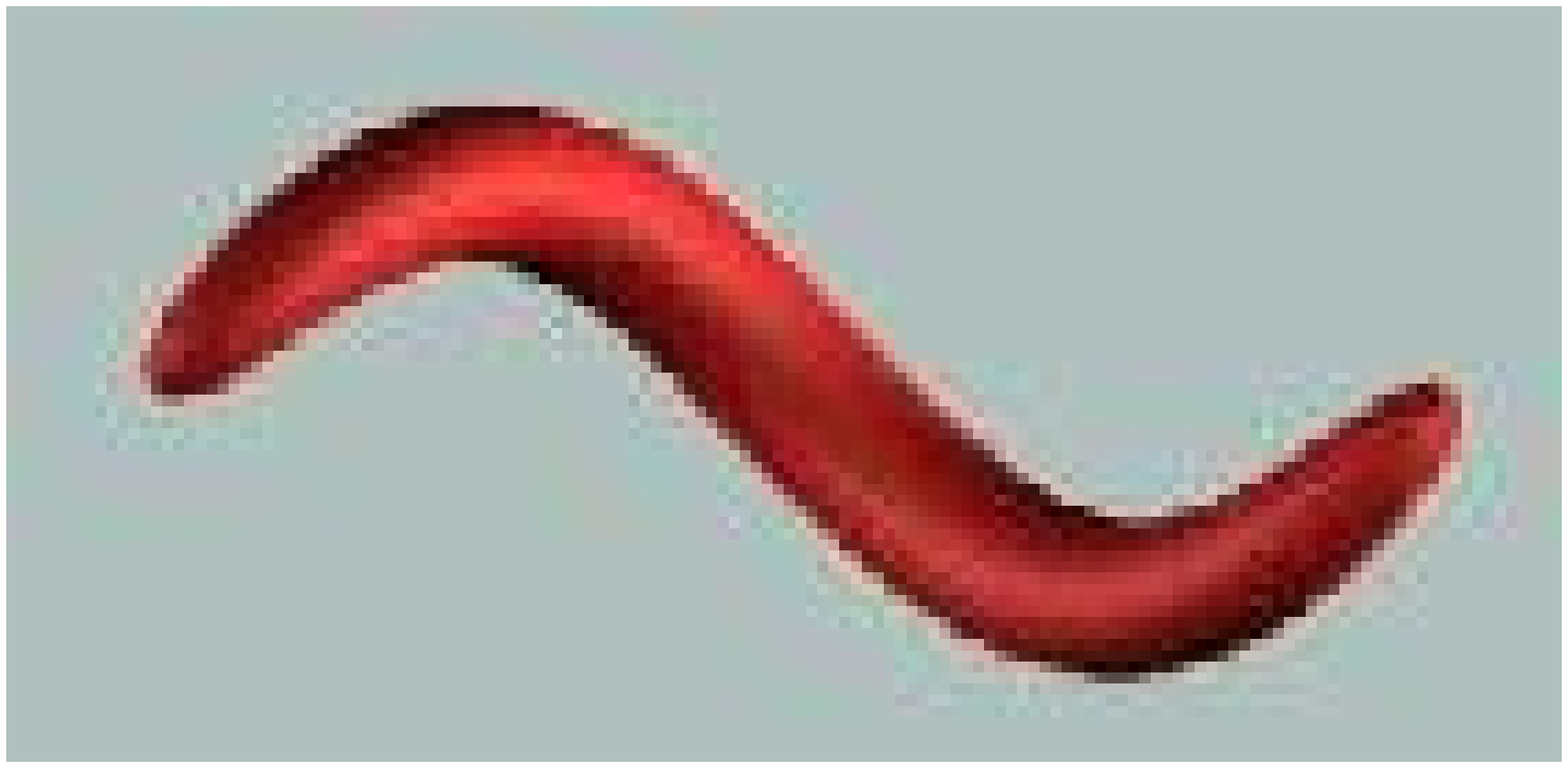,height=1.8cm,width=4.2cm}}}}
\caption{Theoretical example with $\kappa=1$. The Poynting vector is
directed left-to-right.}
\end{figure}
\end{center}

The integral of $\beta_{\mathbb{R}^3}$ over the 3-space for such solutions gives
$$
\int_{\mathbb{R}^3}\beta_{\mathbb{R}^3}=                                       
\kappa E\frac{2\pi\mathcal{L}_o}{c}=
\kappa ET=\pm ET,
$$
where $E=\int_{\mathbb{R}^3}*(Q_{\mu 4}dx^{\mu})=\int_{\mathbb{R}^3}\phi^2d^3v$
is the integral energy of the solution, $T=2\pi\mathcal{L}_o/c$ is
the intrinsically defined time-period, and $\kappa=\pm 1$ accounts for the
two polarizations. Clearly, this integral may be interpreted as
spin-momentum of the solution for one period $T$. Finally, the constant
$\mathcal{L}_o$ may be intrinsically defined by
$$
\mathcal{L}_o=\frac{|A|}{|\delta F|}=\frac{|A^*|}{|\delta *F|}.
$$

\section{Conclusion}
Compare to the classical view, the basic difference of our approach to
relativistically describe electromagnetic field objects consists mainly in the
following steps:
\vskip 0.3cm
	1. These objects are considered as spatially finite and permanently
propagating with the fundamental velocity "c", so their stress-energy-momentum
tensor $Q_{\mu\nu}$ must be {\it null} : $Q_{\mu\nu}Q^{\mu\nu}=0$.

	2. Every electromagnetic field object is a special kind of a physical
object: it is built of two recognizable, time-stable and appropriately
interacting subsystems, formally represented by the two differential 2-forms
$(F,*F)$ on Minkowski space-time.

	3. The {\it null} natuere of these objects:
$Q_{\mu\nu}Q^{\mu\nu}=I_1^2+I_2^2=0$, requires
the two subsystems to carry the {\it same} nonzero energy-momentum:
$I_1=0 \rightarrow F_{\mu\sigma}F^{\mu\sigma}=(*F)_{\mu\sigma}(*F)^{\mu\sigma}$,
and to interact {\it without} interaction energy:
$I_2=0 \rightarrow F_{\mu\sigma}(*F)^{\mu\sigma}=0$.

	4. The two subsystems admit recognizable {\it nonzero} changes,
formally represented by $(\mathbf{d}F,\mathbf{d}*F)$.

	5. The two subsystems live in a permanent dynamical equilibrium
through realizing a local energy-momentum exchange according to the relations:
$$
F^{\mu\nu}(\mathbf{d}F)_{\mu\nu\sigma}=0, \ \
(*F)^{\mu\nu}(\mathbf{d}*F)_{\mu\nu\sigma}=0, \ \
F^{\mu\nu}(\mathbf{d}*F)_{\mu\nu\sigma}=-(*F)^{\mu\nu}\mathbf{d}F_{\mu\nu\sigma},
\ \ \mu<\nu .
$$

	6. The equations, describing the corresponding intrinsic dynamics and
space-time propagation as a whole of an electromagnetic field object through
their nonlinear solutions, represent the understanding that the
$\mathbb{R}^2$-valued 2-vector $\bar{\Omega}=\bar{F}\otimes e_1+\bar{*F}\otimes
e_2$ is $(\mathbf{d},\vee)$ - algebraic and differential symmetry of the
$\mathbb{R}^2$-valued 2-form $\Omega=F\otimes e_1+*F\otimes e_2$ according to
the equations $i^{\vee}_{\bar{\Omega}}\Omega=\mathfrak{C}=0, \
\mathcal{L}^{\vee}_{\bar{\Omega}}\Omega=0$.

	7. The space-time propagation of an electromagnetic field object has
a {\it compatible translational-rotational nature}, and is characterized by a
proportional to the object's integral energy {\it spin momentum},
locally represented by the $F\leftrightarrow *F$ {\it nonzero} local
energy-momentum exchange.

	8. Formally, the existence of non-zero spin momentum is measured by the
Frobenius non-integrability of the two geometric distributions defined
by $\bar{F}=\bar{\zeta}\wedge\bar{A}$ and $\bar{*F}=\bar{\zeta}\wedge\bar{A}^*$.
\vskip 0.3cm
  The last (No.8) step we consider as very suggestive in the following sense.
When a continuous physical system consists of several recognizable interacting
time-stable subsystems, and these subsystems admit natural representation as a
number of geometric distributions and corresponding codistributions, then the
Frobenius integrability relations to be essentially used as local measures of
time stability of the subsystems, and the available curvature forms to be
essentially used as local energy-momentum exchange "communicators" among the
subsytems.  Also, if the whole system admits formal representation as geometric
distribution, in order it to be time-stable, an external local time-like, or
isotropic, symmetry must exist.

\vskip 0.7cm

{\bf References}
 \vskip 0.6cm
[1]. {\bf Stoil Donev, Maria Tashkova},
{\it A nonlinear prerelativistic approach to mathematical representation of
vacuum electromagnetism}, arXiv: hep-th/1303.2808v2

[2]. {\bf H. Lorentz}, {\it Electromagnetic phenomena in a system moving with
any velocity smaller than that of light}, Proc. Acad. Sci., Amsterdam, 1904,
{\bf 6}, (809); {\bf 12} (986)

[3]. {\bf A. Poincare}, {\it Sur la dynamique de l'electron}, Comptes Rendue
1905, {\bf 140} (1504); Rendiconti dei Circole Matematico di Palermo, 1906,
{\bf XXI} (129)

[4]. {\bf A. Einstein}, {\it Zur Elektrodynamik bewegter Korper}, Ann. d. Phys.
1905, {\bf 17} (891)

[5]. {\bf H. Minkowski}, {\it
The Fundamental Equations for Electromagnetic Processes in Moving Bodies},
lecture given at the meeting of the Gottingen Scientific Society, December 21,
1907; english vrsion at: http://www.minkowskiinstitute.org/mip/books/minkowski.html

[6]. {\bf G. Rainich}, {\it Electrodynamics in general Relativity},
Trans.Amer.Math.Soc., {\bf 27} (106-136)

[7]. {\bf W. Misner, J. Wheeler}, {\it Classical Physics as geometry},
Ann.Phys., {\bf 2} (525-603)

[8]. {\bf J. Franca, J. Lopez-Bonilla}, {\it The Algebraic Rainich Conditions},
Progress in Physics, vol.3, July 2007

[9]. {\bf J.L.Synge}, {\it Relativity: The Special Theory},
North-Holland, 1956, Ch.IX, \S\,7.

[10]. {\bf P. Michor}, {\it Topics in Differential Geometry}, AMS, 2008

[11]. {\bf W. Greub}, {\it Multilinear Algebra}, first edition,
Springer-Verlag, 1967

[12]. {\bf M. Forger, C. Paufler, H. Reomer}, {\it The Poisson Bracket for
Poisson Forms in Multisymplectic Field Theory}, arXiv: math-ph/0202043v1

[13]. {\bf C. Godbillon},  {\it Geometrie differentielle et mecanique
analytiqe}, Hermann, Paris (1969)

[14]. {\bf A. Kushner, V. Lychagin, V. Rubtsov,}, {\it Contact Geometry and
Non-linear Differential Equations}, Cambridge University Press 2007

[15]. {\bf Donev, S., Tashkova, M.}, {\it Geometric View on Photon-like
Objects}, arXiv: math-ph/1210.8323 (published as monograph by LAP-Publishing)

\end{document}